\documentclass[pra,showpacs,amsfonts,amsmath,floatfix,onecolumn]{revtex4} 

\usepackage{graphicx}
\usepackage{epstopdf}
\usepackage{amssymb}
\usepackage{amsmath}
\usepackage{xspace}
\usepackage{color}

\begin{document}

\title{Self Trapping Triggered by losses in Cavity QED}
\author{Raul Coto}
\email{rcoto@uc.cl}
\affiliation{Departamento de F\'{i}sica, Pontificia Universidad Cat\'{o}lica de Chile, Casilla 306, Santiago, Chile}
\author{Miguel Orszag}
\affiliation{Departamento de F\'{i}sica, Pontificia Universidad Cat\'{o}lica de Chile, Casilla 306, Santiago, Chile}
\author{Vitalie Eremeev}
\affiliation{Facultad de Ingenier\'ia, Universidad Diego Portales, Santiago, Chile}


\begin{abstract}
In a coupled cavity QED network model, we study the transition from a localized superfluid like state to a delocalized Mott insulator like state, triggered by losses.
Without cavity losses, the transition never takes place.
Further, if one measures the quantum correlations between the polaritons via the negativity, we find a critical cavity damping constant, above which the negativity displays a single peak in the same time region where the transition takes place.
Additionally, we identify two regions in the parameter space, where below the critical damping, oscillations of the initial localized state are observed along with a multipeaked negativity, while above the critical value, the oscillations die out and the transition is witnessed by a neat single peaked negativity.
\end{abstract}

\maketitle

\section{Introduction}
Quantum Phase Transitions(QPT), as opposed to the classical transitions, driven by quantum rather than thermal fluctuations, occur at absolute zero and are induced by the change of some coupling constant or physical parameter. These phenomena have captured the attention of many researchers during the last decade \cite{{Sachdev}, {Carusotto}, {Greiner}, {Stoferle}, {Baumann}, {Osterloh}, {Mekhov}, {wu}}.

 In cavity QED networks, the trade off between the Mott Insulator (MI) and Superfluid (SF) phases have been specially studied. In the present context, these are not truly "phases" in the thermodynamic sense, since in our model, we  deal only with few cavities. However, it has been shown that even small systems, containing a reduced  number of cavities display the superfluid-Mott transition \cite{angelakis2,kim, hartmann}. The importance of this transition is that it enables the system to go from a delocalized state (MI), with the excitations  equally distributed throughout the system, to the state where the excitations move freely (SF), and more importantly, all the excitations could be localized in just one cavity.

In the literature, e.g. \cite{{angelakis2}, {angelakis3}, {kim}}, persists a common conviction that, there is a key parameter that controls this transition, which is the atom-cavity detuning, $\Delta$. In this work we demonstrate the importance of other parameters like damping rate, hopping rate and quantum correlation quantifiers (e.g. Negativity) in detecting and controlling the mentioned phases. 

In many problems of matter-light interactions, it is more convenient to model the atom and cavity field together as a quasi particle called ``Polariton". These polaritons are the dressed states of the Jaynes-Cummings model. However, in the polariton basis, the variation of $\Delta$ leads to a change in the nature of this basis, e.g. in the limit of very large $\Delta$,  the polaritonic state goes to a purely photonic state, with the atoms in the ground state. Therefore, it is meaningless to talk about the SF polaritonic phase in that limit. This problem was addressed  by Irish $\textit{et al.}$ \cite{kim}.
 Nevertheless, a QPT is still possible for small variations of $\Delta$.

Recently, a similar QPT was found in \cite{trapping1,trapping2,trapping3}, by controlling the atom-cavity coupling constant, $\textit{g}$. For a fixed detuning, a variation of $\textit{g}$ leads also to a change in the polaritonic basis, but not for $\Delta=0$, in which case, the states remain always maximally entangled  between light and matter. All these transitions require  an external control parameter, e.g. an external laser acting on each cavity and splitting the atomic levels by Stark shift, which increases $\Delta$.

On the other hand, Quantum Correlations (QC) and their role in Quantum Information is already well known \cite{nielsen,Osterloh,qpt1} and a great effort has been devoted to the understanding of the connection between QC and QPT, see \cite{qpt1} and the references therein. In particular, QC play an important role detecting quantum phase transitions \cite{campbell,phase_transition}. In our case, since we are dealing with qudits, our best candidate for computing the quantum correlations is the Negativity \cite{negativity}.

In this work, we show a critical phenomenon, which is closely related to a QPT, with the losses playing a crucial role.
 We consider as initial a Superfluid like state, and show that it evolves to a Mott Insulator like state during the time evolution of the system, only when the interaction with a reservoir is on. As in many models, we consider that the main source of dissipation originates from the leakage of the cavity photons due to imperfect reflectivity of the cavity mirrors. On the other hand, spontaneous emission will be neglected, assuming long atomic lifetimes, when dealing with the conditions imposed in our cavity QED system.

This paper is organized as follows, we first briefly discuss the mapping of the Jaynes-Cummings-Hubbard model to the Polaritons. Then we study the transition from superfluid like state to Mott insulator like state or self-trapping effect triggered by dissipation using the quantum correlations as a witness. Finally, we analyze the critical damping and discuss the results.

\section{Mapping to Polaritons basis}    

In our model we consider a linear array of N coupled cavities, where each cavity supports a single field mode and contains a single two-level atom. Photons are allowed to hop between neighboring cavities.

 The Hamiltonian of such a system, found in \cite{raul2} and references therein, is given by Eq.($\ref{fullH}$)   

\begin{eqnarray}\label{fullH}
\mathit{H}  &=&  \sum_{j=1}^N \left [ \omega_j^a|e\rangle_j\langle e| + \omega_j^c a_j^{\dagger}a_j + g_j(a_j^{\dagger}|g\rangle_j\langle e|+a_j|e\rangle_j\langle g|)  \right ] \nonumber \\
&+&  \sum_{j=1}^{N-1}J_j[a_j^{\dagger} a_{j+1} +  a_{j+1}^{\dagger}a_j],
\end{eqnarray}
where $|g\rangle$ and $|e\rangle$ are the ground and excited states of the two-level atom with transition frequency $\omega^a$, $a^{\dagger}$($a$) is the creation(annihilation) operator of the cavity mode $\omega^c$, $g$ is the coupling strength between the atom and the cavity mode, $J_j$ is the coupling strength (hopping) between the neighboring cavities and $N$ represents the number of cavities. 
For each cavity, the eigenstates of the first three terms of the Hamiltonian (\ref{fullH})  are the dressed states, which are known as ``polaritons \cite{angelakis2,angelakis3}. These states are given by

\begin{eqnarray}\label{states}
\vert n-\rangle &=& \cos(\theta_n)\vert n,g\rangle -\sin(\theta_n)\vert n-1,e\rangle\nonumber\\
\vert n+\rangle &=& \sin(\theta_n)\vert n,g\rangle + \cos(\theta_n)\vert n-1,e\rangle \nonumber\\
E_{n\pm} &=& \omega^c n+\frac{\Delta}{2}\pm \frac{\sqrt{\Delta^2 +4g^2 n}}{2}  
\end{eqnarray}
with $\Delta=\omega^a-\omega^c$, $\theta_n=\frac{1}{2}\arctan(\frac{g\sqrt{n}}{\Delta/2})$, and $n$ corresponds to the number of photons inside each cavity.

One would expect, since the first three terms of the Hamiltonian (\ref{fullH}) can be written in the polariton basis, that there is a map to rewrite the hopping term (fourth term) in the polaritonic basis. We derive this mapping in a different way as compared to \cite{koch}, but the results are similar,  both being a generalization of the one proposed in \cite{angelakis2}. We found that the creation operator, in the polaritonic basis reads

\begin{equation}\label{adagger}
a^{\dagger} =\sum_{i=1}^n \left (  c_{i+}\mathit{L}_{i+}^{\dagger} + c_{i-}\mathit{L}_{i-}^{\dagger}  \right ) + \sum_{i=2}^n \left ( k_{i\pm}\mathit{L}_{i\pm}^{\dagger} + k_{i\mp}\mathit{L}_{i\mp}^{\dagger}  \right )
\end{equation}

The lowering operator $\mathit{L}_{i+}=\vert (i-1)+\rangle\langle i+\vert$, destroys the state $\vert i+\rangle$. On the other hand, the raising operator $\mathit{L}_{i+}^{\dagger}$ creates the state $\vert i+\rangle$. It is easy to see that with these definitions, all the transitions relative to a state $\vert i \rangle$ will correspond to the upwards and downwards unit jumps.
 The first two terms in (\ref{adagger}) correspond to the transitions between subspaces of the same sign, e.g. from $\vert i-\rangle$ to $\vert(i-1)-\rangle$. The last two terms are related to inter converting the subspace, i.e. from the positive subspace to the negative one, and vice versa.  We notice that the sum for these last terms start at $2$, since  there is no inter converting transition from one to zero. It will be convenient for future notation,  to define each term in the above equation as $\mathit{P}_+^{\dagger}$,$\mathit{P}_-^{\dagger}$,$\mathit{P}_{\pm}^{\dagger}$ and $\mathit{P}_{\mp}^{\dagger}$, in the same order that they appear. The coefficients are found to be

\begin{eqnarray}\label{coefficient}
c_{i+}&=&\sqrt{i}\sin(\theta_i)\sin(\theta_{i-1})+\sqrt{i-1}\cos(\theta_i)\cos(\theta_{i-1}) \hspace*{0.2cm} i>1   \nonumber  \\
c_{i-}&=&\sqrt{i}\cos(\theta_i)\cos(\theta_{i-1})+\sqrt{i-1}\sin(\theta_i)\sin(\theta_{i-1}) \hspace*{0.2cm} i>1   \nonumber \\
k_{i\pm}&=&\sqrt{i}\sin(\theta_i)\cos(\theta_{i-1})-\sqrt{i-1}\cos(\theta_i)\sin(\theta_{i-1}) \hspace*{0.2cm} i\geq 2 \nonumber \\
k_{i\mp}&=&\sqrt{i}\cos(\theta_i)\sin(\theta_{i-1})-\sqrt{i-1}\sin(\theta_i)\cos(\theta_{i-1}) \hspace*{0.2cm} i\geq 2 \nonumber \\
c_{1+}&=&\sin(\theta_1), \hspace*{0.2cm} c_{1-}=\cos(\theta_1)
\end{eqnarray}

Next, we perform some approximations and discuss which of the terms in the above equation can be neglected and under what conditions. 
The inter converting operators, $\mathit{P}_\pm$ and $\mathit{P}\mp$, are the first candidates to be neglected. For only one excitation, this operators vanish, since $k_{\pm}$ start from two excitations. Also, for a higher subspace, it is easy to see that the factor $k_{i}$ decreases considerably. The worst setting is for the second subspace $(i=2)$, if we chose $\theta=\pi/4$, then $k_2\approx 0.21$. If comparing it with other coefficients, i.e. $c_{2+} = c_{2-} \approx 1.21$, we realize that the inter converting operators are not so important during the evolution, and in most of the cases they are completely negligible.

Taking into account the previous discussion, the hopping Hamiltonian can be written as

\begin{equation}\label{hopp2}
\mathit{H}^{hop}=\sum_{j=1}^{N-1} J_j[(\mathit{P}_{(+)j}^{\dagger}+\mathit{P}_{(-)j}^{\dagger})\otimes (\mathit{P}_{(+)j+1}+\mathit{P}_{(-)j+1})+ h.c.]
\end{equation}

Following the idea of reference \cite{angelakis2} and the conditions in \cite{koch}, one can apply the rotating wave approximation in order to eliminate the terms of the type $\mathit{P}_{(+)j}^\dagger \mathit{P}_{(-)j+1}$, since they are fast rotating as compared to $\mathit{P}_{(-)j}^\dagger \mathit{P}_{(-)j+1}$.

Finally, our simplified hopping term in the Hamiltonian, becomes:

\begin{equation}\label{hopp3}
\mathit{H}^{hop}=\sum_{j=1}^{N-1} J_j[\mathit{P}_{(+)j}^{\dagger}\mathit{P}_{(+)j+1} +\mathit{P}_{(-)j}^{\dagger}\mathit{P}_{(-)j+1})+ h.c.]
\end{equation}

Next, if one chooses the initial condition of a particular type, for example $\vert n-\rangle$, then the state $\vert n+\rangle$ will never show up, and we are allowed to consider, throughout the paper, the lower branch only.
It is worth noting that for zero detuning ($\Delta =0$) and restricting the cavities to only the first excited state $\vert 1-\rangle$, we recover our previous results \cite{raul1, raul2}.  However, for finite detuning  and allowing the cavities to have more than one excitation, the dynamics becomes more involved, getting new and interesting results.

\section{Results}
\subsection{Self-Trapping effect witnessed by the Negativity}

To fix the ideas, we imagine starting our system in a Superfluid like state, where all excitations are in a single cavity.
As the system evolves in time, there is a probability that these excitations will be uniformly distributed throughout the system, leading to a Mott insulator like state. This transition is strongly dependent on the detuning.

When the system reaches the Mott insulator like state   $|11\rangle$ , the trapping takes place due to the blockade effect, that is maximized when $\Delta=0$.

\begin{figure}[ht]
\centering
\includegraphics[width=0.45 \textwidth]{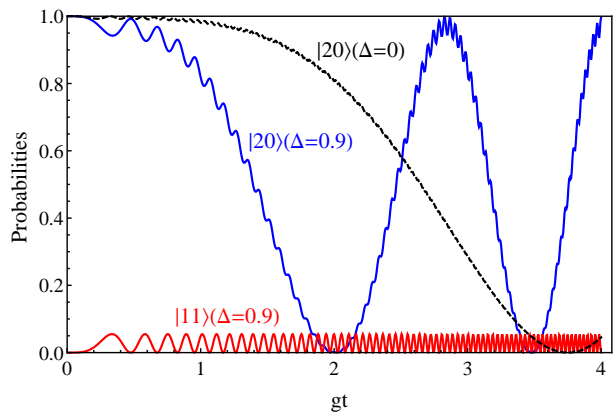}
\caption{For $\Delta=0$ the probability of finding the state $|11\rangle$ is zero, while for $\Delta=0.9g$ is different from zero. Also, the delay in the propagation of the initial state($|20\rangle$) can be shortened by increasing the detuning. Here $\gamma=0$ and $J=0.03g$.} 
\label{fig1}
\end{figure}

In Fig.(\ref{fig1}) we show the time evolution of the probabilities of finding the initial state $|20\rangle$ and $|11\rangle$, when considering only two cavities without losses, for two different cases: with and without detuning. 
On one hand, we notice that for $\Delta=0$, there is delay in the propagation of the initial state to the second cavity, result already discussed in some previous works \cite{trapping1,trapping2,trapping3}.
 Although, in those papers the authors did not study the effect of the detuning to modify (shorten) this delay,  as shown in Fig.(\ref{fig1}), for $\Delta=0.9g$.
We can explain this improvement in the propagation time  based on the fact that when  using the polaritons basis, it is easy to see that an effective hopping shows up, which not only depends on $J$ but also on the coefficient $c_{i-}$ in Eq.(\ref{coefficient}), which increases with the detuning.
 On the other hand, for the case of zero detuning, the probability of finding the state $|11\rangle$ is zero and for   
$\Delta=0.9g$ this probability oscillates with a very small amplitude, as seen from the  Fig.(\ref{fig1}).

In what follows, we will describe a more realistic situation, with each cavity connected to its individual reservoir and consider the same previous initial state, but this time for only the case of zero detuning.
Rather than using the traditional Master Equation approach, we consider the system evolving with a non-Hermitian Hamiltonian, interrupted once in a while by instantaneous quantum jumps, process usually referred to as the ``quantum trajectory" or ``Monte-Carlo wave function method". \cite{QJ}.

\begin{figure}[ht]
\centering
\includegraphics[width=0.45 \textwidth]{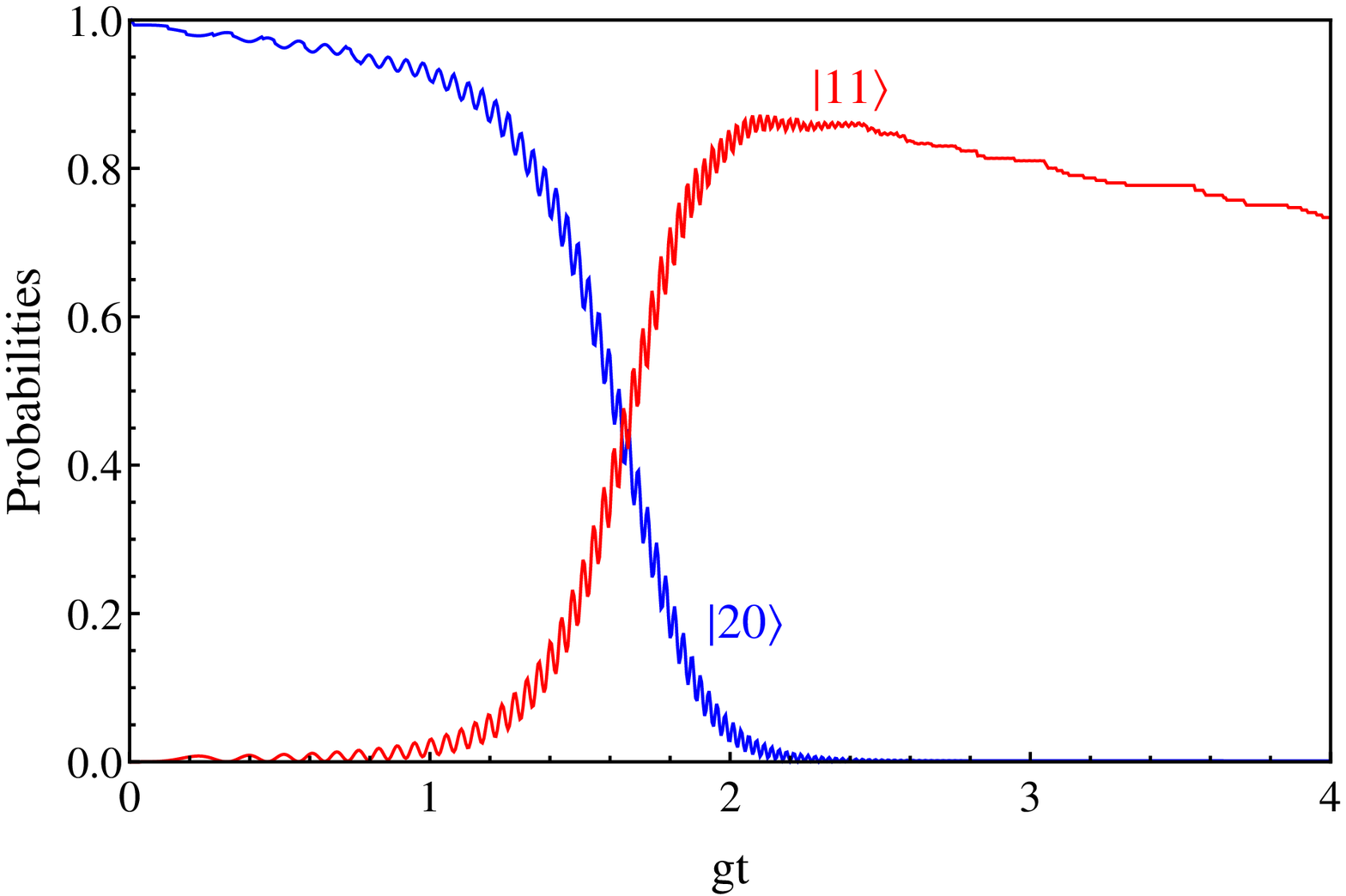}
\includegraphics[width=0.45 \textwidth]{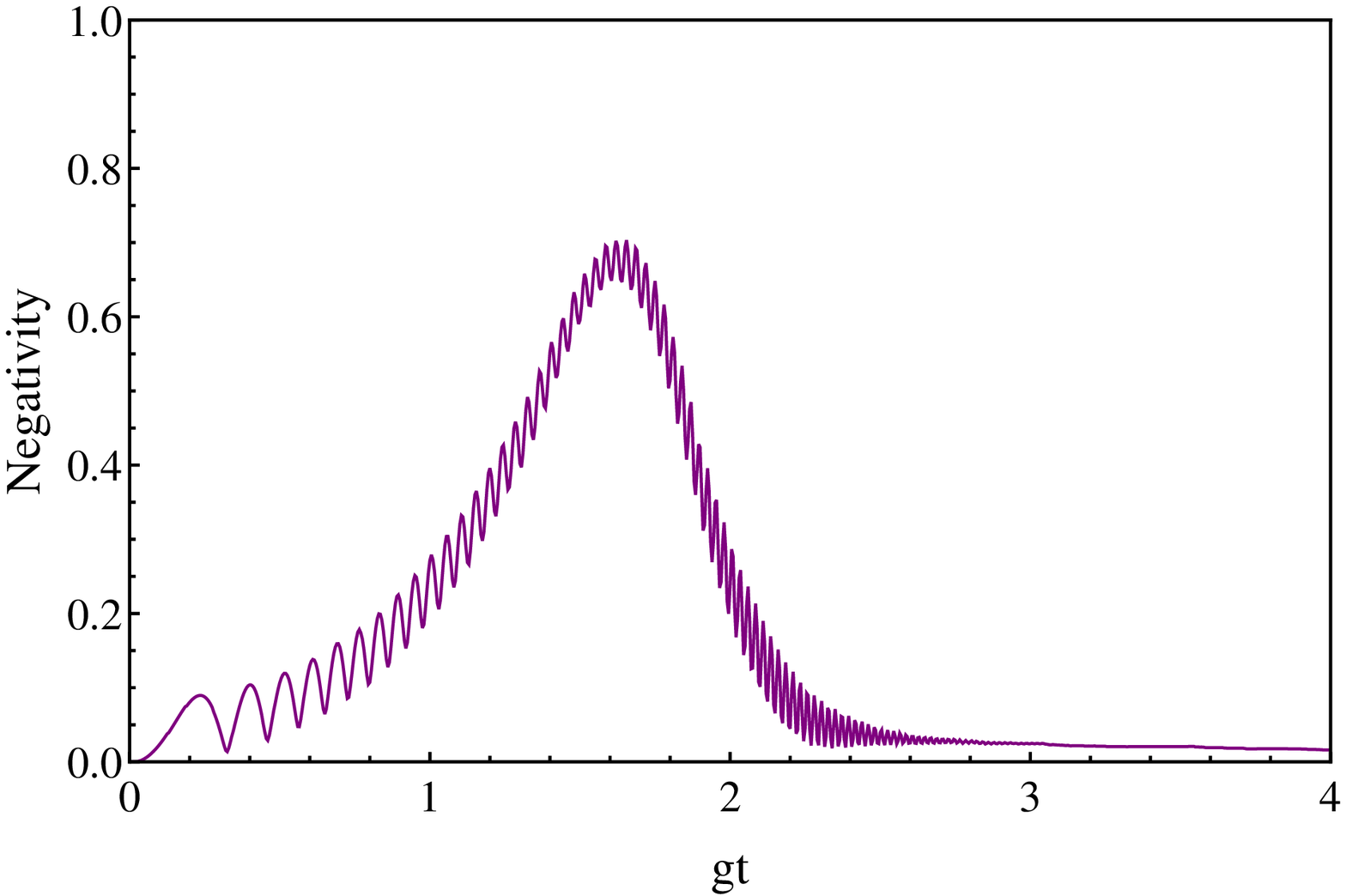}
\caption{When turning on the interaction with the reservoir, a self trapping behavior shows up, where the initial state $|20\rangle$ evolves to the delocalized state $|11\rangle$ and the polariton $|2\rangle$ disappears from the system, indicating that there is no more hopping of the excitations. The Negativity reaches a single maximum in the time region where the transition takes place. Here $\gamma =0.05g$ and $J=0.03g$.} 
\label{fig2}
\end{figure}

 Fig.(\ref{fig2}) shows a self-trapping effect, where the SF like state $|20\rangle$ goes to a MI like state during the time evolution. It is interesting to notice that the presence of the interaction of the reservoir with the cavities triggers the effect, and there is no need of an external control since the system always goes to the same state. 

 For a small loss rate, the polariton $|2\rangle$ oscillates, going back and forth from the first to the second cavity and vice versa,  for certain period of time until a probability associated to the state  $|11\rangle$ starts growing and eventually reaching a slowing decaying plateau. Nevertheless, if we increase the losses to a critical value,  the oscillations vanish and the state $|11\rangle$ appears faster.
  We also show the time evolution of the negativity and observe that at the critical damping, the oscillations of the probability associated to the $|20\rangle$ state die out and there is a single sharp peak in the negativity at the time region where the transition between the states $|20\rangle$ and $|11\rangle$ takes place. 
 We also observe that the critical damping rate depends on the coupling constant between the cavities $J$, as well as 
 the detuning $\Delta$.

\subsection{Criticality with losses }

\begin{figure}[ht]
\centering
\includegraphics[width=0.45 \textwidth]{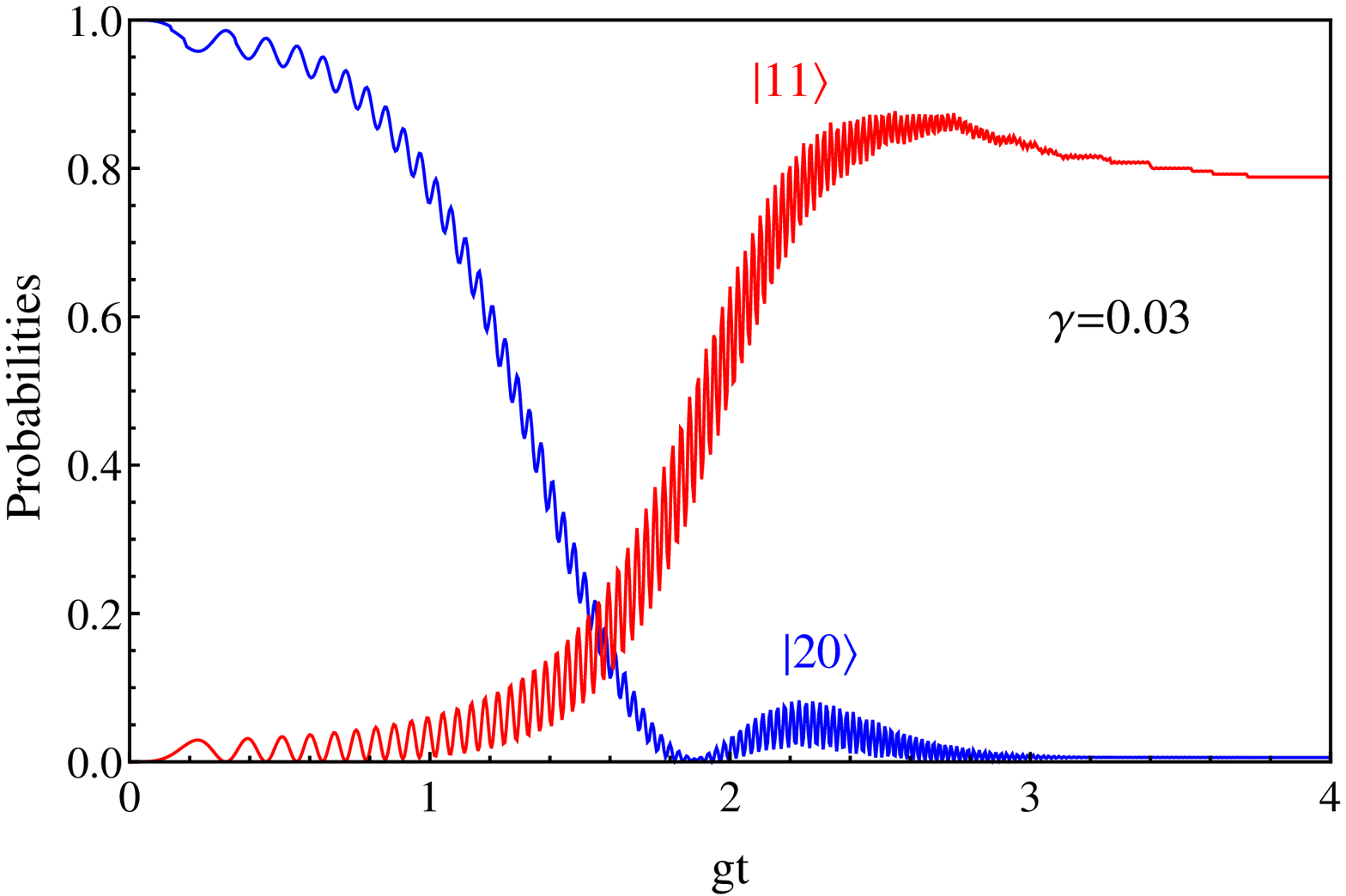} 
\includegraphics[width=0.45 \textwidth]{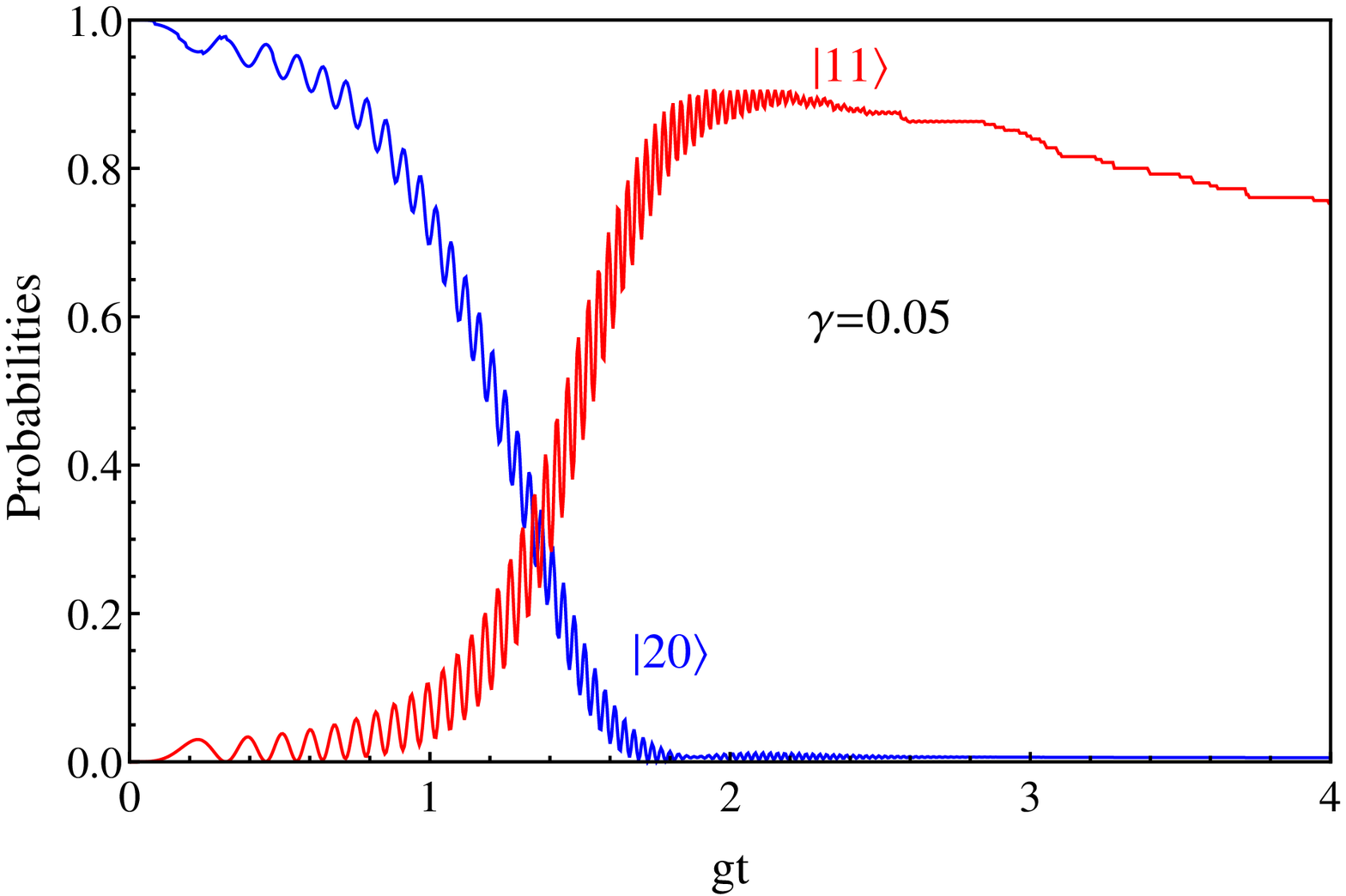}
\includegraphics[width=0.45 \textwidth]{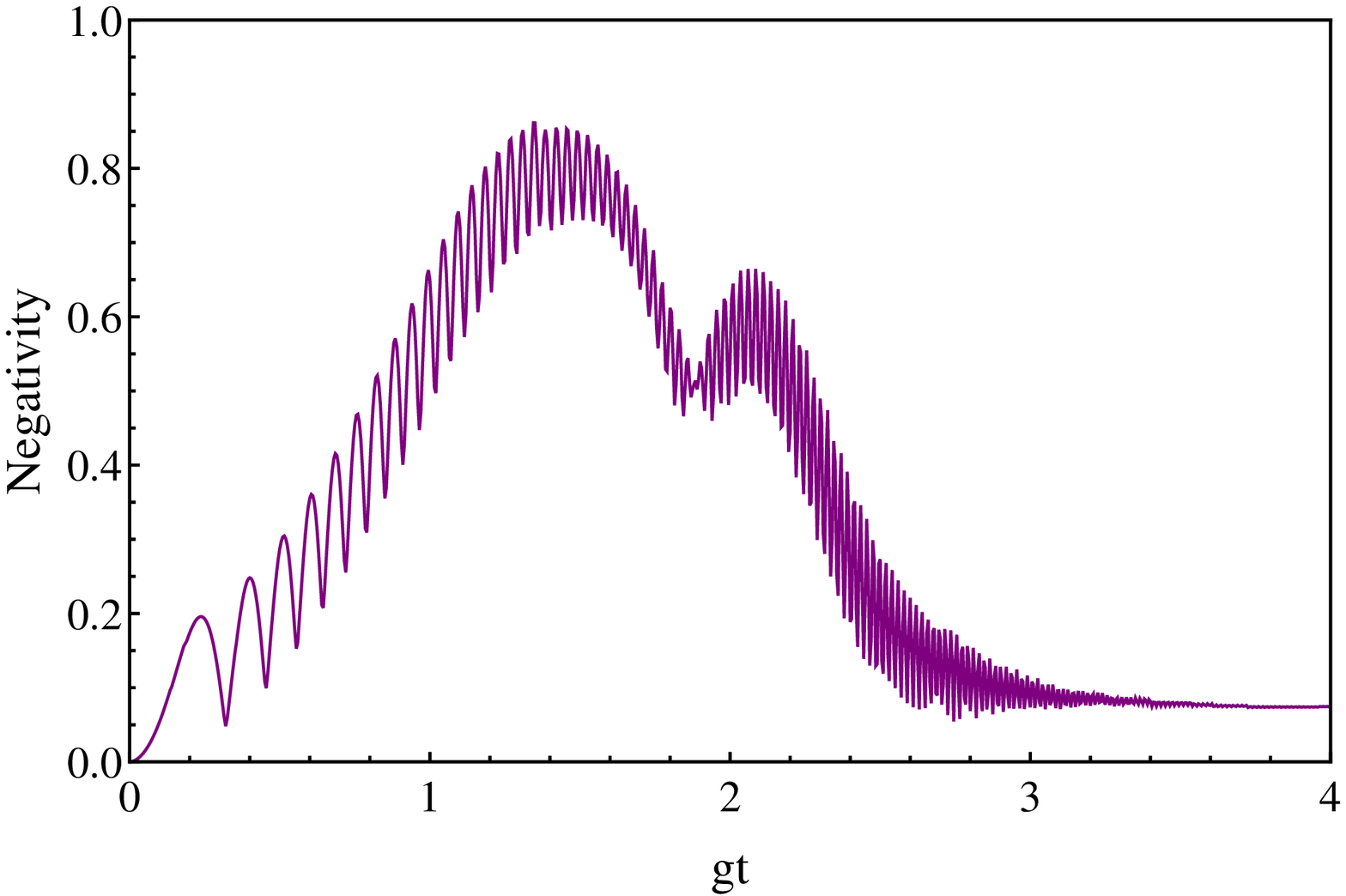}
\includegraphics[width=0.45 \textwidth]{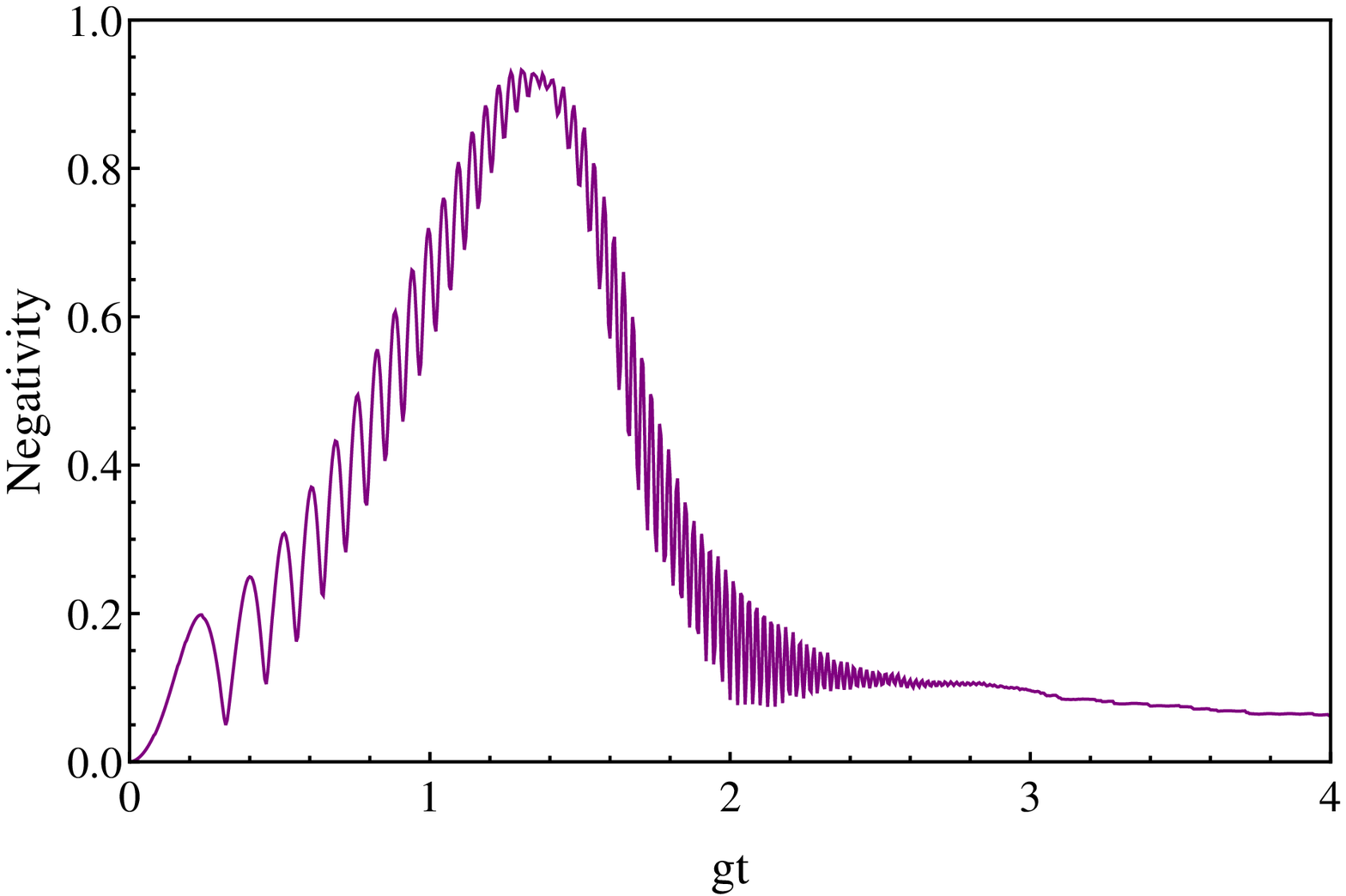}
\caption{ Time evolution of the states $|20\rangle$ (initial one) and $|11\rangle$. From left to right: $\gamma <\gamma_c$ and $\gamma \approx\gamma_c$. The Negativity is a witness to  determine $\gamma_c$. More than one peak implies a backwards motion of the polariton $|2\rangle$. Here $\Delta=0$ and $J=0.06g$.} 
\label{fig3}
\end{figure}

The key feature in this work is the presence of losses. As we pointed out in the previous section, these lead to sudden transitions between two states.

 Furthermore, we are able to find a critical loss rate, $\gamma_c$, for which there is no reflection or oscillation of the probability associated to the polariton $|2\rangle$. This implies a break in the periodicity of the system, where the damping rate is large enough to eliminate the oscillation. 
 
 In the following, we investigate the peculiarities of $\gamma_c$, and notice that it is a function of the coupling strength between the cavities, and also the detuning.
 
In this subsection, we focus on the case $\Delta=0$, with $\gamma_c(J)$ and postpone the discussion on the $\Delta$ dependence to the conclusions. 

As was mentioned before, the negativity can help us to throw some light on the search of criticality.

In Fig.(\ref{fig3}) we present two cases, with $\gamma <\gamma_c$ and $\gamma \approx\gamma_c$. The case $\gamma >\gamma_c$ was discussed in Fig.(\ref{fig2}). 
On one hand, when we are below the critical value, there are oscillations of the initial state, leading to two or more peaks in the  Negativity.
 On the other hand, when going near or above $\gamma_c$, we find a single peak, but as we increase the damping constant, it displaces to the left in time, with a decreasing maximum. Considering these two facts, we numerically estimate the critical damping to satisfy the relation $\gamma_c(J)\approx J$, that corresponds to a negativity with a single peak at its maximum value.
 
\begin{figure}[ht]
\centering
\includegraphics[width=0.45 \textwidth]{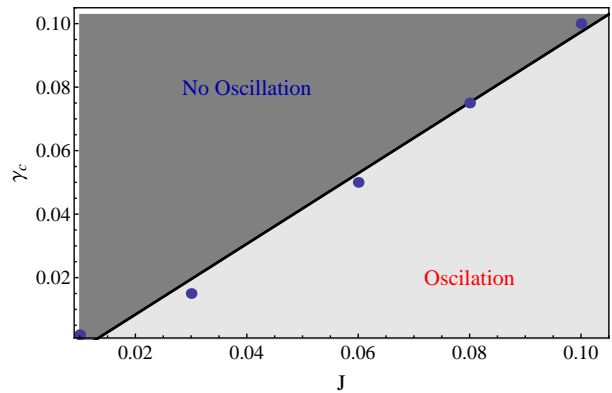}
\caption{The dependence $\gamma_c$ vs $J$ (both in units of $g$) identifying two regions: one with a sudden transition from the state $|20\rangle$ to $|11\rangle$; and the other, where a strong oscillation of the $|20\rangle$ is observed and the transition takes a longer time. Here $\Delta=0$.} 
\label{fig4}
\end{figure}

 Thus, we can clearly identify two regions as shown in Fig.(\ref{fig4}), a lower area that corresponds to an oscillating probability associated to the initial state and the upper region with no oscillations and where the self trapping takes place with a single peaked negativity.

\section{Conclusions}

In this work,  we demonstrated for a system of two coupled cavities with dissipation to individual reservoirs, the occurrence of self-trapping effect, implying a transition from a superfluid to a Mott insulator phase. 

Further, we found a critical damping rate, above which, the initial superfluid states damps out in time while the Mott insulator phase is being created. We find the Negativity to be a very good witness that shows, at the critical damping rate, a single peak at its maximum value in the same time region  where the transition takes place.

We notice that in the lossless case, the initial state oscillates back and forth, that is, the two excitations oscillate between the two cavities and the MI phase $|11\rangle$ will never be reached.
 Hence, we can assert that the true triggering mechanism for this self-trapping effect is the presence of the dissipation in the cavities.

As was discussed above, increasing the detuning also increases the effective hopping, and therefore one would expect a similar dependence of the critical losses with $\Delta$ as with $J$. Even though, this is only true for small variation of $\Delta$, as we increase it, the slope of the straight line shown in Fig.(\ref{fig4}) becomes higher. Moreover, as we increase $\Delta$ further, a rapid oscillation of the Negativity shows up,  which makes it impossible to identify  one single peak. For even larger detuning, say $\Delta = 4g$, the state is almost photonic (there is no entanglement between light and matter), implying that there will be no self-trapping at all, since the atoms are in the ground state and photons can hop freely.

 Furthermore, we studied the behavior of the transition when adding more cavities to the system, up to four. In that case, the MI like state ($\vert 111\rangle$ and $\vert 1111\rangle$ for three and four cavities respectively) shows up at a later time (as compared to the two cavity case), which is to be expected since the system is bigger thus it takes more time to propagate through the array.
 Meanwhile, the decay of the initial state ($\vert 300\rangle$ and $\vert 4000\rangle$) is more abrupt, showing a sign that we are getting closer to a real phase transition in the thermodynamic sense.
This whole analysis was done at zero Temperature. In a near future, we plan to investigate the thermal effects on these transitions.
\cite{Vitalie1,Vitalie2}.

We acknowledge the financial support of the Fondecyt projects no.1140994, the project Conicyt-PIA Anillo ACT-112, "Red de analisis estocastico y aplicaciones", Pontificia Universidad Catolica de Chile, Universidad Diego Portales and Conicyt doctoral fellowships. We also thank Professor M.Loewe for providing extra computer facilities.

\end{document}